\documentclass[journal=jacsat,manuscript=article]{achemso}
\setkeys{acs}{articletitle = true}

\usepackage[version=3]{mhchem} 
\usepackage{amsmath}  
\usepackage{amsfonts} 
\usepackage{subcaption}
\usepackage{graphicx} 
\usepackage{booktabs}
\usepackage{threeparttable}
\usepackage[font=footnotesize]{caption}
\usepackage{verbatim}
\usepackage{xcolor}
\usepackage{soul}
\usepackage{comment}
\usepackage{xr}
\usepackage{pdflscape}
\usepackage{tikz}
\usepackage{geometry}
\usepackage{dsfont}
\usetikzlibrary{matrix,positioning,decorations.pathreplacing} 
\SectionNumbersOn
\mciteErrorOnUnknownfalse
\makeatletter
\newcommand*{\addFileDependency}[1]{
  \typeout{(#1)}
  \@addtofilelist{#1}
  \IfFileExists{#1}{}{\typeout{No file #1.}}
}
\makeatother

\usepackage{blindtext}
\usepackage{graphicx}
\usepackage{mhchem}

\newcommand{\ha}{\mathrm{Ha}^{-1}}

\newcommand*{\myexternaldocument}[1]{%
    \externaldocument{#1}%
    \addFileDependency{#1.tex}%
    \addFileDependency{#1.aux}%
}

\myexternaldocument{SupportingInfo}

\title{The Design of New Practical Constraints in Auxiliary-Field Quantum Monte Carlo}

\author{John L. Weber}
\email{jlw2245@columbia.edu}
\author{Hung Vuong}
\affiliation{Department of Chemistry, Columbia University, 3000 Broadway, New York, NY, 10027}
\author{Richard A. Friesner}
\author{David R. Reichman}
\email{drr2103@columbia.edu}
\date{May 2023}

\begin{document}

\begin{abstract}
We formulate and characterize a new constraint for Auxiliary Field Quantum Monte Carlo (AFQMC) applicable for general fermionic systems, which allows for the accumulation of phase in the random walk but disallows walkers with a magnitude of phase greater than $\pi$ with respect to the trial wave function. For short imaginary times, before walkers accumulate sizable phase values, this approach is equivalent to exact free projection, allowing one to observe the accumulation of bias associated with the constraint and thus estimate its magnitude \textit{a priori}. We demonstrate the stability of this constraint over arbitrary imaginary times and system sizes, highlighting the removal of noise due to the fermionic sign problem. Benchmark total energies for a variety of weakly and strongly correlated molecular systems reveal a distinct bias with respect to standard phaseless AFQMC, with a comparative increase in accuracy given a sufficient quality of trial wave function for the set of studied cases. We then take this constraint, referred to as linecut (lc-) AFQMC, and systematically release it (lcR-AFQMC), providing a route to obtain a smooth bridge between constrained AFQMC and the exact free projection results.
\end{abstract}
\maketitle
\section{Introduction}
Projector Quantum Monte Carlo (PQMC) algorithms formally provide a polynomial-scaling route to chemical accuracy for a variety of molecular and solid state systems. However, as the exact projection of an initial guess onto the ground state necessarily encounters a sign or phase problem for general fermionic systems, the key to the extended success of PQMC has been the formulation of constraints which effectively remove the sign problem while retaining a high level of accuracy. Phaseless Auxiliary-Field Quantum Monte Carlo (ph-AFQMC) has provided a prominent example of such a constrained method in the basis of non-orthogonal Slater determinants.\cite{shee2023potentially} It does so by exactly restricting the propagation of walkers to a single gauge in the complex plane via a ``force bias," defined in relation to a trial wave function. This cancels the accumulation of phase from the propagator to first order in the imaginary timestep, $\Delta\tau$. In practice, one must additionally remove the origin from the calculation to avoid the sign problem. There are many possible ways to do so, as originally described by Zhang \textit{et. al.} in the paper that first introduced the phaseless approximation.\cite{zhang2005quantum} Zhang \textit{et. al.} found that the most accurate and stable algorithms for model systems were obtained when using a cosine projection to scale the weight of each walker according to the stepwise growth of phase, namely
\begin{equation}
    W_w = |W_w| \times \max(\cos(\Delta\theta), 0),
\end{equation}
where $W_w$ is the weight of the walker indexed by $w$, and $\Delta\theta$ refers to the stepwise change of phase with regard to the trial wave function, $\theta=\arg(\langle \Phi_T | \Phi_w \rangle)$. In the case of an exact trial wave function this procedure provides exact results, however for an inexact trial wave function, this results in a bias, commonly referred to as the ``phaseless bias." 

The behavior of the phaseless bias is difficult to quantify, and thus the vast majority of studies to date focus on benchmarking ph-AFQMC with different trials in order to devise strategies to obtain increasingly accurate benchmark results on difficult systems at minimal cost. Such approaches have been largely successful, with calculations using multi-reference trials performing well on many transition metal complexes for which single reference methods such as coupled cluster can fail.\cite{rudshteyn2020predicting,shee2019singlet,williams2020direct,neugebauer2023toward,rudshteyn2021calculation}

The most popular strategy to reduce the cost of AFQMC, and thus render it scalable to larger, more interesting systems, is to provide more affordable ways to converge the trial wave function, e.g. by using selected CI wave functions\cite{mahajan2021taming,mahajan2022selected} or by updating a single determinant trial until self consistency is obtained.\cite{he2019finite,qin2023self} Recently, coupled cluster (CC) trials were used with the assistance of a quantum computer \cite{huggins2022unbiasing}. Additionally, one can devise systematically convergable approximations\cite{weber2022localized,malone2018overcoming} or sampling strategies\cite{shee2017chemical} that reduce the cost of running AFQMC with a fixed trial, which can, in some cases, take advantage of an enhanced cancellation of error for energy differences. While these approaches are useful in practice, there still exist many cases where the performance of ph-AFQMC is worse than expected, given the nature of the problem and the quality of the trial. For example, recent work by Lee \textit{et. al.} pointed out the failure of ph-AFQMC with an unrestricted Hartree Fock (UHF) trial to adequately describe open shell atoms.\cite{lee2022twenty} Indeed, we have found that even given a small CASSCF trial wave function describing the valence S and P orbitals (Xe4o), ph-AFQMC does not attain chemical accuracy for any of the open shell main group atoms (Fig. \ref{figure:wedgeatoms}). While numerically exact energies may be obtained with a converged trial or via non-constrained (free projection) AFQMC, these results highlight the limitations of exclusively using previous benchmarks to determine when a trial is sufficient for ph-AFQMC.  Additionally, many single reference transition metal problems still appear to require a large number of determinants in the trial wave function to adequately converge the total energy for ph-AFQMC, which can result in significant increases in cost for each ph-AFQMC calculation. The prospect of modifying the phaseless constraint in the hopes of obtaining less biased results given a fixed trial thus remains very enticing.

Two decades after the invention of ph-AFQMC we now have sufficient computational power to rigorously test alternatives to the cosine projection on benchmark sets, and efforts to do so have already emerged in the literature. For example, Sukurma et. al. \cite{sukurma2023benchmark} recently tested a re-formulation of the cosine projection in which they delay projection until evaluating observables, instead removing the origin by killing walkers if their total phase grows beyond $\frac{\pi}{2}$. They found qualitatively similar yet statistically distinct results to ph-AFQMC on a small benchmark set of small main group molecules.

Another approach to systematically improve the accuracy of ph-AFQMC is to interface it in some way with the exact non-constrained free projection AFQMC (fp-AFQMC), known as constraint release.\cite{shi2013symmetry,mahajan2021taming} For ph-AFQMC, it is necessary to reformulate the importance sampling procedure upon releasing the constraint, as left in the constrained form the force bias effectively removes walkers which violate the phaseless constraint by reducing the weight to zero even in the absence of explicitly removing them upon crossing the origin. This procedure is typically unstable, and one is thus restricted to fully removing the constraint at once, which necessarily reduces the applicability of constraint release techniques in AFQMC. 

Recently, Xiao \textit{et. al.} introduced a framework which bypasses this instability by initiating a non-constrained Markov chain Monte Carlo calculation with a population of walkers from a constrained ph-AFQMC random walk which was subsequently back-propagated.\cite{xiao2023interfacing} With sufficiently long imaginary times, as estimated by the variance in energy decreasing beyond a certain threshold, they were able to obtain results starting from a ph-AFQMC calculation which are exact within error bars. Moreover, by freezing a portion of the back-propagated trajectory within the Markov chain simulation, they were able to lessen the severity of the observed sign problem at the cost of retaining some bias.

In constrast to this study, we choose to forego the goal of eliminating the phase entirely, and report on a new class of ``phaseful'' constraints. We do not make use of importance sampling with the force bias, and we allow arbitrary phase accumulation up to $|\theta| = \alpha \pi$. This effectively breaks the symmetry of the random walk in the complex plane while simultaneously removing the origin, allowing for a retention of signal with a bounded noise amplitude. For the open shell atoms, we find that this constraint is stable up to $\alpha = 1.0$, contrary to the common expectation of the emergence of an exponential phase problem for $\alpha > 0.5$ (see Fig. \ref{figure:wedgeatoms}). For $\alpha > 1.0$ our approach reduces to free projection. This formalism introduced here enables a novel means to release the constraint and thus allows for a systematic partial sampling of the unbiased electronic structure problem. 

This paper is organized as follows: We first provide a brief overview of the details of the phaseless AFQMC algorithm and introduce our new approach in this context. We then present results for the open shell atoms when scanning $\alpha$, showing stability up to $\alpha = 1.0$ and a consistent improvement in accuracy versus ph-AFQMC. We then explore the accuracy of the $\alpha = 1.0$ constraint, which we call ``linecut'' (lc-) AFQMC, for a series of model molecular systems for which exact total energies are available, including \ce{H_4} dissociation, \ce{N_2} dissociation, \ce{C_2}, and benzene. We test for size consistency using an increasing number of \ce{N_2} molecules, and observe similar behavior to ph-AFQMC when using a large timestep. We additionally show consistent stability as the system size increases, reproducing a phaseless (exact) calculation for \ce{Ni(Cp)2} in the cc-pVTZ basis within statistics. Finally, we present an algorithm for the partial release of the linecut constraint (lcR-AFQMC), asymptotically reproducing the unbiased free projection. Partially released calculations are shown to allow for the convergence of AFQMC calculations for difficult systems given a fixed trial. We conclude with an outlook on the practical use of our approach within the context of AFQMC calculations.

\section{Theory}
Here we give a brief overview of the formalism of phaseless AFQMC, followed by a description of the proposed constraint. For a more in depth review of ph-AFQMC, we suggest some recent reviews.\cite{shi2021some,motta2018ab,shee2023potentially} In PQMC, initial states $|\Phi_i \rangle$ which are non-orthogonal with respect to the exact ground state will recover the exact ground state $| \Phi_0 \rangle$ upon projection in imaginary time $\tau$, 
\begin{equation}
    \lim_{\tau \to \infty} e^{-\tau \hat{H}} | \Phi_i \rangle = | \Phi_0 \rangle,
\end{equation}
where $\hat{H}$ denotes the electronic Hamiltonian, $\hat{H} = \hat{H}_1 + \hat{H}_2 = \sum_{pq} h_{pq} c^{\dag}_p c_q + \frac{1}{2} \sum_{pqrs} \langle pq|rs \rangle c^{\dag}_p c^{\dag}_q c_s c_r$. Time evolution is formulated in terms of a series of finite steps in imaginary time, $e^{-\tau\hat{H}} = (e^{-\Delta\tau\hat{H}})^n$, each of which are then separated into one- and two- body terms via a symmetric Suzuki-Trotter decomposition,
\begin{equation}
    e^{-\Delta\tau(\hat{H}_1 + \hat{H}_2)} \simeq e^{-\frac{\Delta\tau\hat{H}_1}{2}}e^{-\frac{\Delta\tau\hat{H}_2}{2}} e^{-\frac{\Delta\tau\hat{H}_1}{2}} + \mathcal{O}(\Delta\tau^3).
\end{equation}
If we write the electronic two-body operator as a sum of one-body operators squared, $\langle pq|rs \rangle = \sum_{\alpha} L_{pr,\alpha}L_{qs,\alpha}$ via exact diagonalization or approximate methods such as density fitting or modified Cholesky decomposition,\cite{purwanto2011assessing} we can then use the Hubbard-Stratonovich identity to convert the two-body operators into a multi-dimensional integral over a set of fluctuating ``auxiliary-fields'' x$_{\alpha}$,
\begin{equation}
    e^{-\frac{\Delta\tau}{2}(\sum_{\alpha} L_{\alpha}^2)} = \prod_{\alpha}{\int_{-\infty}^{\infty} \frac{1}{\sqrt{2\pi}} e^{-\frac{x_{\alpha}^2}{2}} e^{\sqrt{-\Delta\tau}x_{\alpha}L_{\alpha}}dx_{\alpha}} + \mathcal{O}(\Delta\tau^2).
    \label{hubstrat}
\end{equation}
It is this multi-dimensional integral on which we perform Monte Carlo sampling
\begin{equation}
 |\Phi(\tau + \Delta\tau) \rangle=\prod_{\alpha}\int_{-\infty}^{\infty} \frac{1}{2\pi} e^{-\frac{x_{\alpha}^2}{2}} e^{\sqrt{-\Delta\tau} x_{\alpha}L_{\alpha}}dx_{\alpha}|\Phi(\tau)\rangle= \int d\mathbf{x} P(\mathbf{x}) \hat{B}(\mathbf{x}) |\Phi(\tau)\rangle,
 \label{endMC}
\end{equation}
where $\mathbf{x}$ is the vector of auxiliary fields. In AFQMC, this simulation can be formulated as a branching, open-ended ensemble of random walkers $w$ over the manifold of Slater determinants, each represented by a single Slater determinant $|\Phi_{\tau,w}\rangle$ and corresponding weight $W_{\tau,w}$. Each walker is propagated forward at a given time $\tau$ by $\hat{B}(\mathbf{x}_{\tau,w})$, with the space of auxiliary fields \textbf{x} being sampled from the Gaussian probability defined in Eq. \ref{hubstrat}. As $\hat{B}(\mathbf{x}_{\tau,w})$ is represented by purely one body operators, propagation respects the Thouless theorem and produces another Slater determinant, thereby maintaining the anti-symmetry of the wave function.\cite{thouless1960stability} 

To reduce the necessary sampling, the weights are updated according to the ratio of the new overlap with the trial to the previous overlap
\begin{equation}
    |\Phi_{\tau + \Delta\tau,w} \rangle = \hat{B}(\mathbf{x}_{\tau,w}) |\Phi_{\tau,w} \rangle,
\end{equation}
\begin{equation}
    W_{\tau + \Delta\tau,w} e^{i\theta_{\tau + \Delta\tau,w}} = \frac{\langle \Phi_T  | \Phi_{\tau + \Delta\tau,w} \rangle}{\langle \Phi_T |\Phi_{\tau,w} \rangle}W_{\tau,w} e^{i\theta_{\tau,w}}.
    \label{eq:weight_update}
\end{equation}
The AFQMC representation of the wave function in this importance sampling framework is thus given by 
\begin{equation}
    |\Phi\rangle = \sum_w W_w \frac{|\Phi_w\rangle}{\langle\Phi_T|\Phi_w\rangle}.
\end{equation}

So far, the algorithm presented is formally exact and is referred to as free projection (fp-) AFQMC; however, the un-constrained fermionic phase problem leads to an exponential decrease in the signal-to-noise ratio with respect to imaginary time duration as the walkers are propagated. In the phaseless constraint, we exactly shift the auxiliary fields by a complex ``force bias'',\cite{zhang2003quantum,rom1998shifted}

\begin{equation}
 \prod_{\alpha}\int_{-\infty}^{\infty} \frac{1}{2\pi} e^{-\frac{x_{\alpha}^2}{2}} e^{\sqrt{-\Delta\tau} x_{\alpha}L_{\alpha}}dx_{\alpha}|\Phi(\tau)\rangle= 
 \prod_{\alpha}\int_{-\infty}^{\infty} \frac{1}{2\pi} e^{-\frac{(x_{\alpha}-\bar{x}_{\alpha})^2}{2}} e^{\sqrt{-\Delta\tau} (x_{\alpha}-\bar{x}_{\alpha})L_{\alpha}}dx_{\alpha}|\Phi(\tau)\rangle,
 \label{projector_with_FB}
\end{equation}
where $\bar{x}_{\alpha}$ is chosen to cancel the accumulation of phase with respect to the trial to first order,
\begin{equation}
\bar{x}_{\alpha} = \frac{\langle \Phi_T  | \hat{L}_{\alpha} | \Phi_{w} \rangle}{\langle \Phi_T |\Phi_{\tau,w} \rangle}.
\end{equation}
This modifies the weight $W_w$ by a factor $e^{x_{\alpha}\bar{x}_{\alpha}-\frac{\bar{x}_{\alpha}^2}{2}}$. In practice we must additionally remove the origin with respect to the trial, which is accomplished via the cosine projection, which projects the weight back onto the real axis after each step and removes walkers which become negative, namely 
\begin{equation}
    W_w = W_w \times \max(0, \cos(\Delta\theta)). 
    \label{phaseless}
\end{equation}

In our new constraint, we do not make use of the force bias, instead beginning from the free projection projector defined in Eq. \ref{endMC}. We update the weights as in Eq. \ref{eq:weight_update}, keeping track of the overall phase accumulated. As soon as the magnitude of the phase of a walker exceeds $\alpha\pi$, the walker weight is updated to zero and thus annihilated,
\begin{equation}
    W_w = \left\{
    \begin{array}{lr}
        W_w, & \text{if } |\theta_w| < \alpha\pi \\
        0, & \text{if } |\theta_w| > \alpha\pi
    \end{array}
    \right.
    \label{eq:linecut}
    .
\end{equation}
This constraint with $\alpha = 1.0$ is what we henceforth refer to as linecut (lc-) AFQMC.

\section{Results}
\subsection{Open Shell Atoms}
\begin{figure}
    \centering
    \includegraphics[width=10cm]{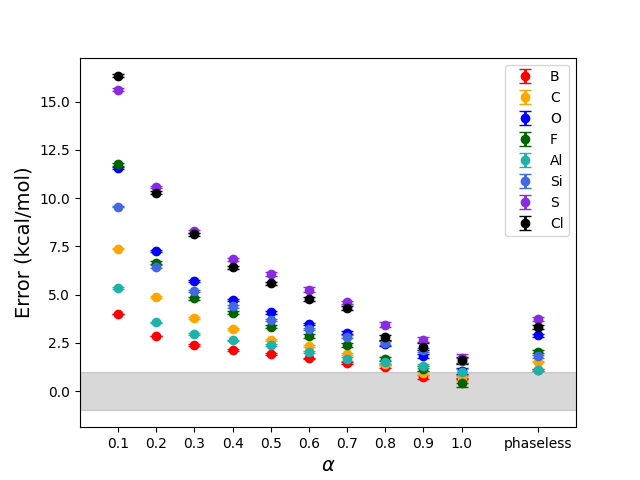} 
    \caption{Systematic convergence with $\alpha$ for the open shell atoms of AFQMC with the constraint as specified in Eq. \ref{eq:linecut}, versus ph-AFQMC. All calculations use CASSCF trials including the valence S and P orbitals in the active space. The grey bar indicates chemical accuracy, or 1 kcal/mol.
    } 
    \label{figure:wedgeatoms}
\end{figure} 

Recently, Lee \textit{et. al.} reported substantial biases for the set of open shell atoms using ph-AFQMC with a UHF trial,\cite{lee2022twenty} making them ideal small candidates for the testing of new constraints. We thus use these atoms with small RCAS trials to test the performance and stability of the new approach with increasing $\alpha$. We find the observed bias decreases monotonically as a function of $\alpha$ for all atoms (Fig. \ref{figure:wedgeatoms}). As expected, the variance increases as a function of $\alpha$, with $\alpha \leq 0.5$ exhibiting variance on par with the phaseless constraint, albeit producing much less accurate results due to the removal of the force bias, which thus allows for an accumulation of phase due to propagation. The most interesting cases are those with $\alpha$ above 0.5, where walkers begin to accrue a partial sign. In these cases, while the variance does increase, it does not exponentially increase, as one would expect for a full sign problem (and, indeed, as we see with $\alpha > 1$). Instead, the variance is limited to $\simeq 2-4$ times that found with the phaseless constraint, while the accuracy continues to improve. In the most extreme case, $\alpha = 1$, we see consistently improved results versus ph-AFQMC, bordering on and in some cases obtaining chemical accuracy. For the remainder of the study, we focus on this case, which we refer to as ``linecut'' (lc-) AFQMC.

\subsection{The \ce{C2} Dimer}
\begin{figure}
    \centering
    \includegraphics[width=10cm]{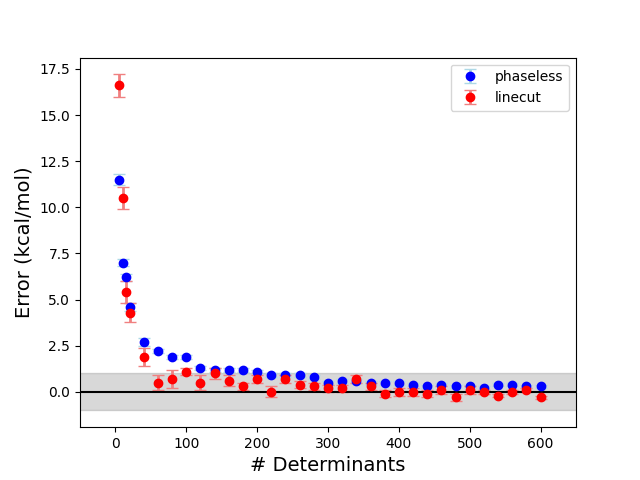} 
    \caption{Systematic convergence of lc- and ph- AFQMC for \ce{C2} with an increasing number of determinants retained in a heat bath CI trial (active space of 8e16o). In each case the determinants of a specified number with the highest weight in the wave function are kept. 
    } 
    \label{figure:C2scan}
\end{figure}

Recently we reported the convergence of ph-AFQMC for the \ce{C2} dimer total energy with respect to the number of determinants used in a small selected CI trial.\cite{shee2022potentially} To test for convergence with respect to the quality of the trial, we perform the same calculations using lc-AFQMC. Initially we observe worse performance from lc-AFQMC when using up to 10 of the highest weighted determinants in a SCI wave function, but lc-AFQMC quickly overtakes that of ph-AFQMC, reaching chemical accuracy at $\simeq$100 determinants, as opposed to the 220 determinants required for ph-AFQMC (Fig. \ref{figure:C2scan}) to reach a comparable level of accuracy. Both methods converge to the exact ground state energy within statistical error bars as the number of determinants increase. It is interesting that with the lower quality (as measured by a small number of determinants) trials, ph-AFQMC outperforms lc-AFQMC, highlighting that although lc-AFQMC converges to chemical accuracy at a faster rate, the behavior of the observed bias is distinct and in some cases can produce less accurate results than standard ph-AFQMC. 

\begin{figure}[H]
    \centering
    \includegraphics[width=10cm]{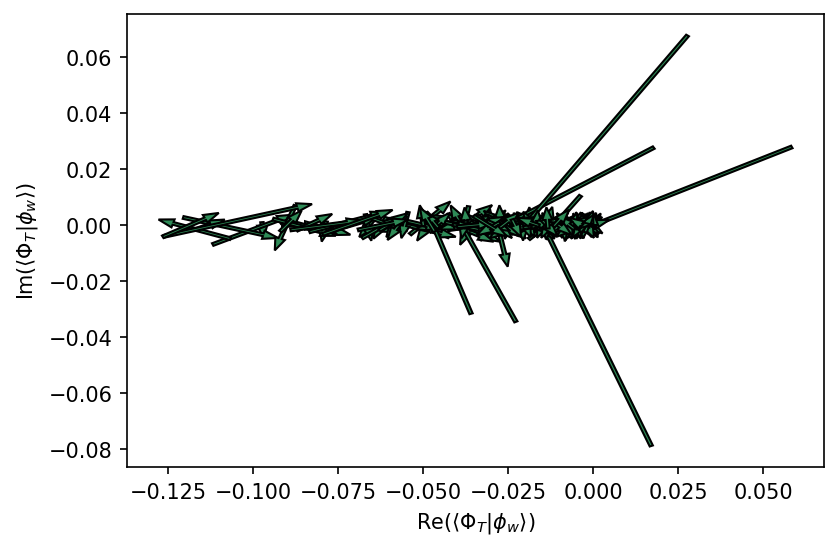} 
    \caption{A sample of walker steps from a lc-AFQMC trajectory which cross the negative real axis, resulting in the elimination of the walker by setting the weight to zero. The tails of the arrows represent the overlap of the walker with the trial before the propagation step, while the head represents the overlap after. lc-AFQMC is stable due to the annihilation of walkers whose overlaps with respect to the trial wave function cross the negative real axis, including the origin, from either side. 
    } 
    \label{figure:linecut_killed}
\end{figure}

\begin{figure}[H]
    \centering
    \includegraphics[width=10cm]{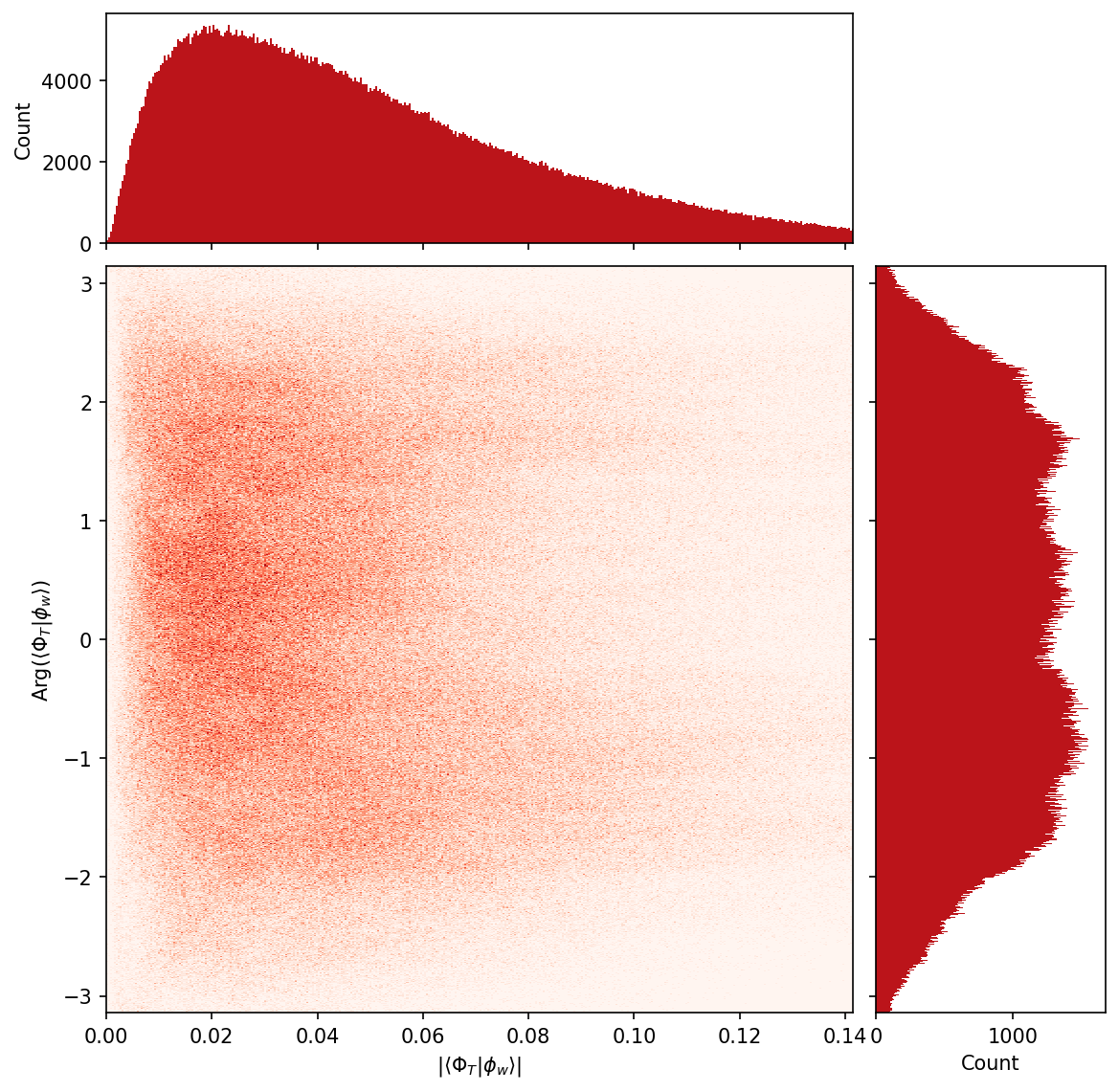} 
    \caption{Distribution of overlaps in the complex plane with respect to the trial wave function for 256 walkers from $100$ to $120\ \ha$ in a lc-AFQMC calculation. Note the effect of the linecut on the space surrounding the negative real axis.
    } 
    \label{figure:lccomplex}
\end{figure}

\begin{figure}[H]
    \centering
    \includegraphics[width=10cm]{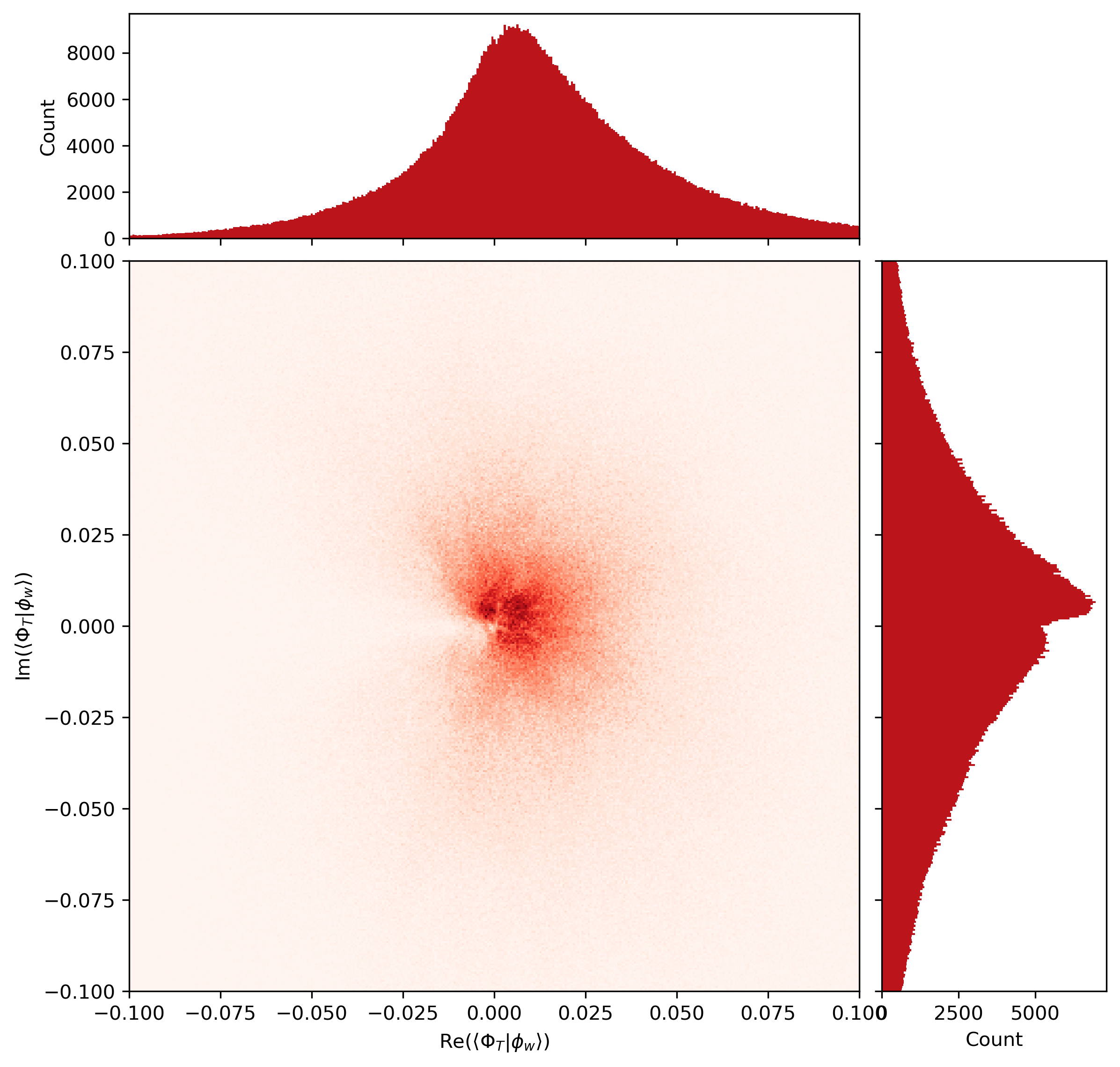} 
    \caption{Distribution of the norm and phase of overlaps with respect to the trial wave function for 256 walkers from $100$ to $120\ \ha$ in a lc-AFQMC calculation.
    } 
    \label{figure:lcnormvsphase}
\end{figure}

The stability of lc-AFQMC for values of $\alpha$ up to $\alpha = 1$ is somewhat surprising. To shed liught on this behavior, we investigate the population of walkers throughout the simulation for the case of the carbon dimer. Upon equilibration, we find that the removal of walkers as they cross the negative real axis (i.e. going from $\theta = \pm(\pi - \epsilon)$ to $\theta = \pm(\pi + \epsilon)$, Fig. \ref{figure:linecut_killed}), along with the use of a finite time step, results in a smooth population density reduction which goes to zero on the negative real axis (see Fig. \ref{figure:lccomplex}-\ref{figure:lcnormvsphase}). Allowing phase accumulation for $|\theta| > \frac{\pi}{2}$ does result in phase cancellation, which, along with the lack of importance sampling, is the cause of the observed increase in variance. However, breaking the symmetry of the random walk in the complex plane enables the signal to be measured above the noise. The largest signal emanates from walkers with overlap closest to one with respect to the trial. As there is no phase cancellation at this point, a more extreme dependence on the quality of the trial can result. Crucially, the lc-AFQMC removes the origin from the calculation and thus removes the sign problem even while allowing phase accumulation. These calculations show no increase in noise as a function of $\tau$ once the phase distribution has equilibrated, and have been run stably for upwards of $1200\ \ha$.

\subsection{Bond Breaking Model Systems}
\begin{figure}
    \centering
    \includegraphics[width=10cm]{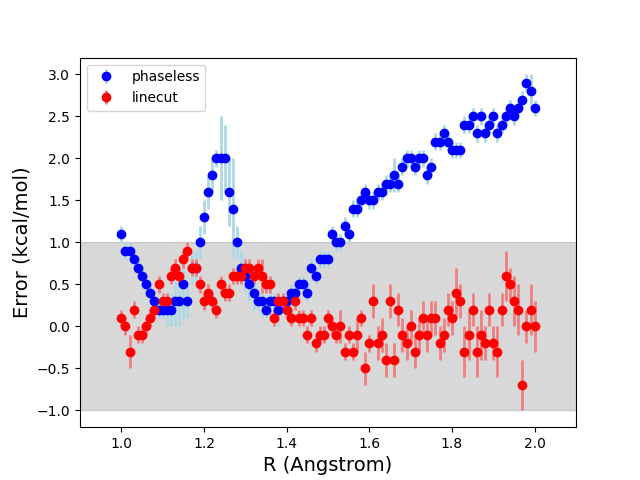} 
    \caption{Error (kcal/mol) of both phaseless and linecut AFQMC for the dissociation of the \ce{H4} molecule in the STO3G basis using a UHF trial. The grey bar indicates chemical accuracy, or $<1$ kcal/mol error.
    } 
    \label{figure:H4_STO3G}
\end{figure}

\begin{figure}
    \centering
    \includegraphics[width=10cm]{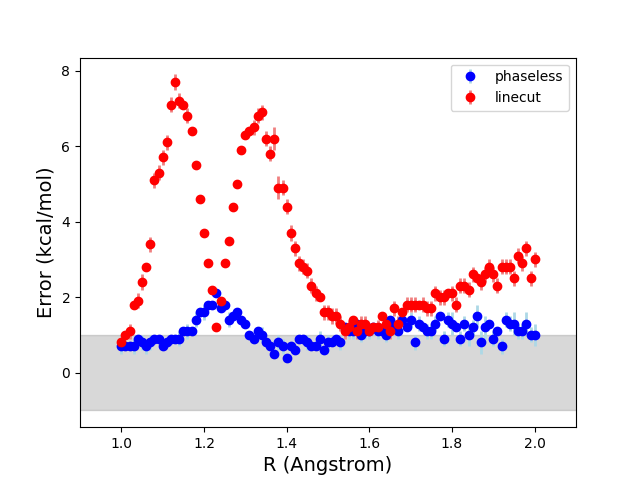} 
    \caption{Error (kcal/mol) of both phaseless and linecut AFQMC for the dissociation of the \ce{H4} molecule in the cc-pVQZ basis using a UHF trial. The grey bar indicates chemical accuracy, or $<1$ kcal/mol error.
    } 
    \label{figure:H4_ccpvqz}
\end{figure}

\begin{figure}
    \centering
    \includegraphics[width=10cm]{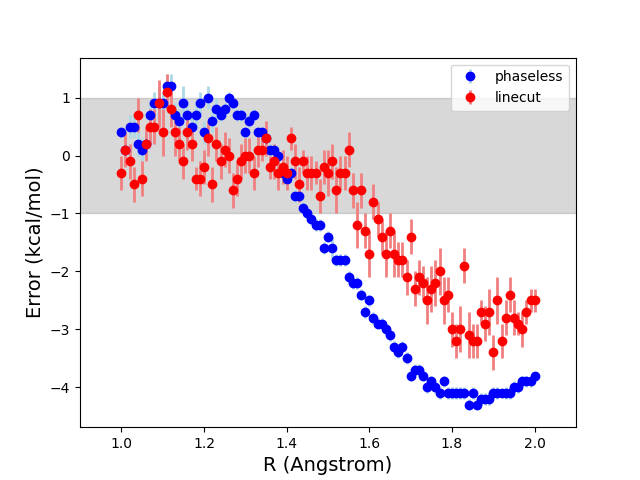} 
    \caption{Error (kcal/mol) of both phaseless and linecut AFQMC for the dissociation of the \ce{N2} molecule in the STO3G basis using a UHF trial. The grey bar indicates chemical accuracy, or $<1$ kcal/mol error.
    } 
    \label{figure:N2_STO3G}
\end{figure}

Two widely studied model systems for strong correlation are the dissociation of the \ce{H4} and \ce{N2} molecules, for which ph-AFQMC results have been recently reported.\cite{lee2022twenty} Following this study, we investigate the performance of lc-AFQMC versus ph-AFQMC for these systems using UHF trials. For \ce{H4} in the STO3G basis, which is thought to accentuate the effects of strong correlation, lc-AFQMC obtains chemical accuracy for all bond distances, including obtaining the correct dissociation limit, albeit with some non-monotonic noise fluctuations as a function of $R$ (Fig. \ref{figure:H4_STO3G}). This behavior is not maintained, however, when increasing the basis set size to cc-pVQZ, where lc-AFQMC performs poorly, especially in the regions of strong correlation surrounding the equal distance points (Fig. \ref{figure:H4_ccpvqz}). Interestingly, the equal distance point at $R = 1.23$ is one of the only cases where lc-AFQMC performs better than ph-AFQMC, and the errors of the two constraints oppose each other. This may suggest that the two constraints complement each other, with lc-AFQMC performing well for systems dominated by static correlation (as in the equal distance point) while ph-AFQMC performs well for systems dominated by dynamic correlation, although it is impossible to generalize such observations with this limited dataset. Neither ph-AFQMC nor lc-AFQMC converges to the correct dissociation limit. To test if the linecut bias is remedied in the larger basis by the use of more accurate trial functions, we investigate \ce{H4} in the cc-pVQZ basis at $R = 1.13$, where we had observed the maximum error of lc-AFQMC, using a 4e4o RCAS trial. Using only 2 determinants, the lc-AFQMC bias is reduced to $3.0\pm0.9$ kcal/mol versus $4.0\pm0.3$ kcal/mol for ph-AFQMC. For \ce{N2}, lc-AFQMC provides a less dramatic improvement over ph-AFQMC in the minimal basis, but is more accurate for nearly all bond lengths (Fig. \ref{figure:N2_STO3G}).


\begin{figure}
    \centering
    \includegraphics[width=10cm]{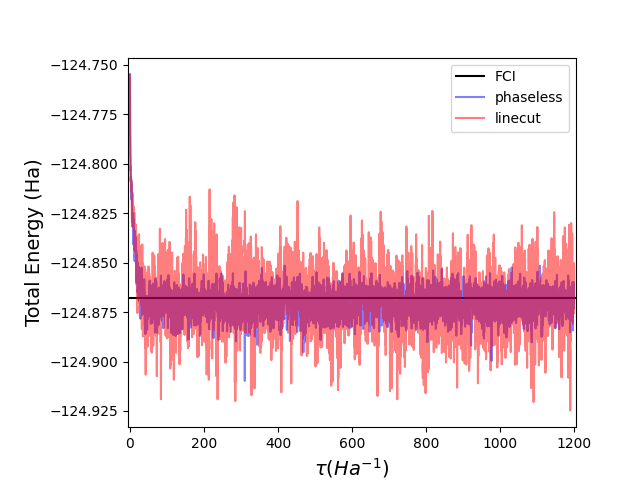} 
    \caption{Trajectories of ph-AFQMC and lc-AFQMC using the exact trial for \ce{N2} at $1.5\ \mathrm{\AA}$. ph-AFQMC is in blue, with lc-AFQMC in a semi-transparent red overlaid on top.
    } 
    \label{figure:ph_vs_lc_traj}
\end{figure}

To showcase the stability of the linecut constraint in imaginary time, we provide a plot of the \ce{N2} trajectories for both ph- and lc- AFQMC at $R = 1.5$ using the UHF trial; both are stable up to $1200\ \ha$ and exhibit similar equilibration times. However lc-AFQMC converges to within statistical error of the exact FCI value ($0.3 \pm 0.4$ kcal/mol off), whereas ph-AFQMC is biased by $1.4 \pm 0.1$ kcal/mol (Fig. \ref{figure:ph_vs_lc_traj}). 

Benchmark total energies for benzene in the cc-pVDZ basis are often additionally used to evaluate high level electronic structure methods for performance in systems dominated by dynamic correlation.\cite{eriksen2020ground} Here, we ran lc-AFQMC with a single determinant trial (RHF), and obtain a total energy of $-862.1(1.0)$ Ha, within statistical error of the accepted exact energy $(-863)$, significantly improving upon the RHF/ph-AFQMC value of $-866.1(3)$ reported in Ref. \citenum{lee2020performance}.

To test whether the linecut constraint stability decreases as a function of system size (and thus a more severe sign problem), we revisit a calculation from a previous study for which ph-AFQMC obtains chemical accuracy versus experiment, namely the case of Nickelocene in the cc-pVTZ-DK basis using an RCAS trial.\cite{rudshteyn2021calculation} We again observe a $\simeq$4x increase in variance versus phaseless that nevertheless remains stable over the course of the simulation. Here, lc-AFQMC ($-2812.432(4)$ Ha) agrees with ph-AFQMC in total energy to within statistical error ($-2812.429(1)$ Ha).

\subsection{Linecut Release}
Due to the lack of importance sampling in the current implementation of lc-AFQMC, it is straightforward to implement a systematic release of the constraint to the free projection limit by relaxing the boundary conditions. In this protocol, walkers are instead tagged to be annihilated in $X$ steps upon violation of the constraint, allowing a partial sampling of the region surrounding the origin. At $X=0$, this is fully constrained lc-AFQMC, whereas at $X=\infty$ the walkers are freely projected and a full sign problem is encountered. For intermediate $X$, one can partially remove the bias while damping the resulting noise from the sign problem. We have demonstrated this with \ce{C2} using a 10-determinant trial (Table \ref{tab:release}). This procedure is similar in spirit to release-node diffusion Monte Carlo. \cite{ceperley1984quantum}

\begin{table}[]
    \centering
    \begin{tabular}{c|c c}
$X$	& Error (kcal/mol)	& st err \\
\hline
0	& 8.4	& 0.6 \\
20 & 	4.0 & 	0.6 \\
40 & 	2.6 & 	0.8 \\
60 & 	0.7 & 	1.5 
    \end{tabular}
    \caption{Systematic release of the linecut constraint for \ce{C2} using a trial with the ten highest weight CI determinants in the wave function. All calculations use 1920 walkers and propagate for $300\ \ha$.}
    \label{tab:release}
\end{table}

\begin{figure}
    \centering
    \includegraphics[width=10cm]{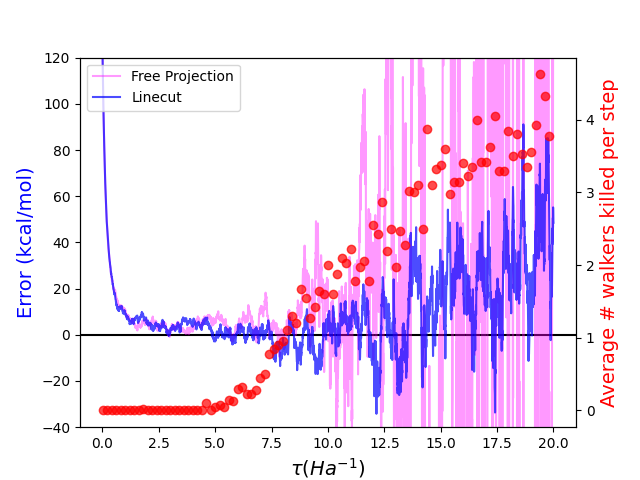} 
    \caption{Error of AFQMC trajectories versus exact energy for FeO (def2-SVP, 864 determinants) using linecut and free projection AFQMC. The two are statistically equivalent (although run with different random walks) until the linecut constraint activates at around $5\ \textrm{Ha}^{-1}$, at which point fp-AFQMC encounters the sign problem and loses signal, and lc-AFQMC incurs a large bias. Both trajectories use 5000 walkers.
    } 
    \label{figure:nkill}
\end{figure}

Whereas partially releasing the linecut constraint over long trajectories does remove the bias while still retaining a reasonable signal to noise ratio for \ce{C2}, we expect that releasing the constraint will be more difficult in more strongly correlated systems where the sign problem is more severe. As an example of such a case, we test the FeO dimer, which requires $\mathcal{O}(10^5)$ or more determinants in a full active space selected CI trial for ph-AFQMC to converge to chemical accuracy. Testing fully constrained lc-AFQMC, the convergence to the exact result is significantly slower than ph-AFQMC, similar to the trials with the lowest number of determinants for \ce{C2} (Table \ref{tab:FeO_reg_error}). However, closer analysis of the trajectories reveals another potential advantage of the linecut constraint; as we neglect to perform any importance sampling, lc-AFQMC is rigorously equivalent to fp-AFQMC until the first walker is killed. By plotting the average number of walkers removed by the constraint each step (Fig. \ref{figure:nkill}), we can observe the nearly exact (and up until this point, sign problem free) trajectory become biased. As many free projection calculations make use of energies obtained at around 5 Ha$^{-1}$, the point where the linecut constraint begins removing walkers, any significant deviation past this provides a clear \textit{a priori} estimate of the bias due to the linecut constraint, and thus the quality of the trial.

\begin{table}[]
    \centering
    \begin{tabular}{c|c c c c}
CI $\%$	&	ph error	&	ph st error	&	lc error	&	lc st error\\
\hline									
90	&	5.5	&	0.2	&	88.8	&	1.8	\\
95	&	3.2	&	0.1	&	69.8	&	1.6	\\
96	&	3.0	&	0.1	&	62.4	&	2.2	\\
97	&	2.4	&	0.1	&	45.7	&	1.3	\\
98	&	2.1	&	0.1	&	22.2	&	0.8	\\
99	&	1.3	&	0.1	&	7.7	&	2.4	
    \end{tabular}
    \caption{Errors (kcal/mol) of phaseless and linecut AFQMC at long imaginary times for \ce{FeO} (def2-SVP) using different numbers of determinants from a selected CI trial with a full active space. lc-AFQMC performs significantly worse than ph-AFQMC. Note that due to cost we only ran lc-AFQMC at 99$\%$ CI weight (26,071 determinants) for $60\ \textrm{Ha}^{-1}$, hence the higher reported statistical error. }
    \label{tab:FeO_reg_error}
\end{table}

\begin{figure}
    \centering
    \includegraphics[width=10cm]{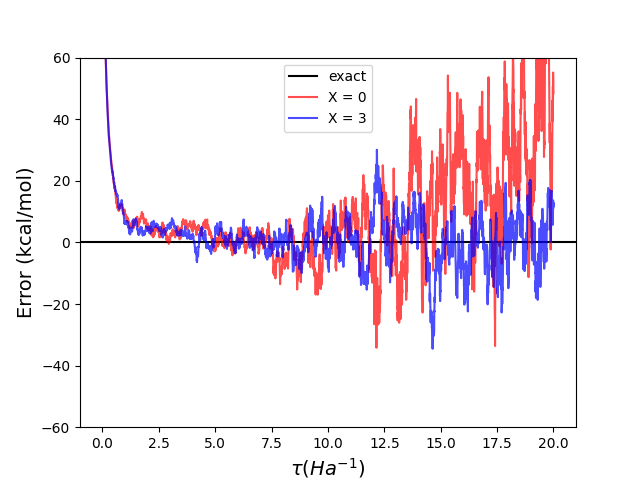} 
    \caption{Error of AFQMC trajectories versus exact energy for \ce{FeO} (def2-SVP, 864 determinants) for $X = 0$ and $X$ = 3, where we observe the minimum of statistical error. By partially releasing the constraint we delay the onset of both the linecut bias and sign problem noise is delayed for long enough in imaginary time to obtain time-averaged results within chemical accuracy of the exact answer.
    } 
    \label{figure:FeO_traj}
\end{figure}

For particularly problematic cases such as FeO, we propose taking further advantage of this relationship between linecut release AFQMC (lcR-AFQMC) and free-projection AFQMC (fp-AFQMC) by obtaining energies at short imaginary times. By slowly releasing the linecut constraint, one can effectively delay the onset of the linecut bias while minimizing the noise growth of the sign problem, allowing for the estimation of energies over a small portion of the short time trajectory, rather than the typical method of running $\mathcal{O}(10^5)$ trajectories to estimate the energy at one timestep as in free projection. Choosing the equilibration time is additionally straightforward, as one can choose the point directly prior to the activation of the linecut constraint. We find for such simulations of \ce{FeO}, we can systematically reduce the bias with increasing $X$, and additionally reduce the statistical error, as the overall change in the trajectory that is averaged over is additionally reduced with the bias (Fig. \ref{figure:FeO_traj}). At a certain $X$, the statistical error begins increasing again due to the sign problem, and lcR-AFQMC no longer provides a significant statistical advantage (Table \ref{tab:FeO_release}).

\begin{table}[]
    \centering
    \begin{tabular}{|c|c |c|}
    \hline
$X$	&	Error (kcal/mol)	&	standard error	\\
\hline					
0	&	5.7	&	3.5	\\
1	&	3.9	&	2.9	\\
2	&	0.7	&	1.3	\\
3	&	0.4	&	0.9	\\
4	&	1.5	&	1.5	\\
5	&	0.3	&	1.6	\\
\hline
FP	&	-57.6	&	51.4	\\
phaseless	&	2.7	&	0.3	\\
\hline
    \end{tabular}
    \caption{Systematic release of the linecut constraint for FeO using 864 determinants in the trial. All calculations use 5000 walkers and are averaged over $4$ to $20\ \textrm{Ha}^{-1}$. Note the initial decrease in statistical error with $X$, due to delaying the onset of the linecut bias, followed by an increase at $X=4$ due to the sign problem.}
    \label{tab:FeO_release}
\end{table}

\section{Conclusions} \label{conclusions}
We have formulated a new constraint for use within the AFQMC framework, which allows phase accumulation up to $|\theta| = \pi$ and retains stability for long timescales. Tests on a variety of small systems demonstrate comparable accuracy with qualitatively distinct behavior versus ph-AFQMC, with some complimentary advantages for lc-AFQMC. When sizable errors do arise in lc-AFQMC, they may be remedied by the use of systematically more sophisticated trials, akin to standard practice in ph-AFQMC and are easily identified by increased variance. 

The stability of the lc-AFQMC approach suggests that the majority of the phase problem of AFQMC manifests from walkers crossing the origin in the complex plane, rather than generic accumulation of phase with respect to the trial wave function. This suggests the possibility of the design of alternative constraints that avoid origin crossing, and thus may serve as a first step in the creation of useful new constraints.

With respect to the release of the linecut constraint, we emphasize that such calculations are asymptotically unstable, as they allow a finite sampling of the exponential sign problem. Nevertheless this technique appears to be numerically useful for systems in which the trial is already qualitatively correct, as in the case of \ce{C2}. Additionally, monitoring the trajectory as the constraint is activated allows for a systematic internal estimate of the constraint bias, which we view as a significant advantage over alternate algorithms. Similar procedures for constraint release may not be as effective within ph-AFQMC due to the use of importance sampling (i.e. the force bias), which effectively removes walkers that violate the constraint indirectly. In this framework, one is restricted to either fully constrained or fully unconstrained calculations, which lessens the applicability, although it has been effectively used in studying lattice models.\cite{shi2013symmetry} A recent work\cite{xiao2023interfacing} describes a method to remove this instability within ph-AFQMC through the use back propagation followed by a non-constrained Markov chain Monte Carlo simulation. It is of course in principle possible to develop an importance sampling method akin to the standard force bias which respects the linecut constraint and thus reduces the variance within lc-AFQMC; in order for this to be useful, however, whatever savings obtained due to reduced sampling would have to outweigh the increased cost and complexity of partially releasing the constraint, as well as the added cost for computing the importance sampling function. If the variance of lc-AFQMC without importance sampling remains 2-4x as large as that of ph-AFQMC, this would result in approximately 4-16x the sampling necessary to obtain comparable error bars. However, it is likely that with increasing system size the increase in variance will eventually become a high enough detriment to necessitate importance sampling.

While the results presented here are promising, we stress that benchmarking on larger datasets is paramount to determine the ultimate utility of lc-AFQMC for generating benchmark energies at lower cost than ph-AFQMC. Indeed, our results even on a small benchmark set do not suggest uniform improvements over ph-AFQMC. We thus suggest that the use of lc-AFQMC in this way is restricted to trials for which systematic extrapolation to the unbiased limit is possible, or to trials which do not show significant deviation upon the onset of the constraint at short imaginary times. To this end, lcR-AFQMC may prove a useful tool by assisting in converging results for difficult systems or in helping diagnose insufficient trials. 

Although our conclusions with regards to relative accuracy versus ph-AFQMC are somewhat preliminary, the stability of lc-AFQMC represents a new class of constrained AFQMC which has significant utility independent from direct usage. In particular, the observed distinct and in some cases directly opposing trends of lc-AFQMC versus ph-AFQMC (as in the dissociation curve of \ce{H4}), suggest a fertile research pathway towards understanding the behavior of bias in constrained AFQMC. By changing the constraint, lc-AFQMC provides another parameter beyond simply increasing the complexity of the trial wave function with which to study the relationship between the trial and constraint bias, and will likely lead to more insight into the performance and occasional failures of ph-AFQMC. More in-depth studies of both the phaseless and linecut biases are necessary to determine if one can correlate the relative accuracy with specific features of the trial function. In addition to more fundamental research such as this, it is possible to merely use lc-AFQMC as a measure of trial quality for general ph-AFQMC simulations by monitoring the change in energy upon initialization of the constraint for short trajectories. This can additionally provide clear internal metrics with which to study the relationship between trial quality and biases in constrained AFQMC.  More extensive benchmark studies which may help answer these and related questions will be pursued in future work.

\section{Acknowledgements}
JLW wishes to thank Shiwei Zhang for the many insightful discussions on the nature of the phaseless constraint in AFQMC. This research used resources of the Oak Ridge Leadership Computing Facility at the Oak Ridge National Laboratory, which is supported by the Office of Science of the U.S. Department of Energy under Contract DE-AC05-00OR22725. JLW was funded in part by the Columbia Center for Computational Electrochemistry (CCCE).

\section{Associated Content}
The supporting information includes: 
\begin{itemize}
\item Additional details of AFQMC calculations
\item Discussion of size consistency at finite timesteps
\end{itemize}
\newpage

\bibliography{References} 

\end{document}


\maketitle
\begin{suppinfo}
\beginsupplement
\section{Details of AFQMC calculations}
All ph-AFQMC and lc-AFQMC calculations were run on our custom AFQMC code with 0.005 $Ha^{-1}$ timesteps unless otherwise noted. Trials were generated using PySCF, and the integrals were decomposed using the modified Cholesky Decomposition algorithm with a threshold of 1e-6. Selected CI trials were generated using Dice interfaced with PySCF. All calculations made use of the ``comb'' population control algorithm upon every energy evaluation, which occurs every 20 steps.

We tested for the convergence of lc-AFQMC with respect to timestep and walker population on $H_4$ in the cc-pVQZ basis at $R = 1.13$, where we saw the largest error out of all linecut calculations. Neither doubling the walker population nor halving the timestep led to a change in the result outside of statistics, suggesting that the simulation is converged with respect to these parameters.

\section{Size Consistency and Stability}
It has been recently pointed out that ph-AFQMC for a fixed and finite timestep is not size-consistent. This was attributed to the cosine projection, and posited that the removal of this from the constraint could result in a size consistent method even for finite time step size.\cite{lee2022twenty} To this end, we tested a series of \ce{N2} molecules for size consistency using a fairly large timestep of $0.05\ \ha$ (Fig. \ref{figure:size_consistency}). While the better performance of lc-AFQMC is evident vs ph-AFQMC, it still exhibits growth of correlation energy as the number of \ce{N2} molecules is increased, suggesting that the removal of the cosine projection is not the only requirement for a size consistent constraint prior to the limit $\Delta\tau \to 0$ being taken. It is necessary to again note that this does not preclude general size consistency for constrained AFQMC, which is rigorously size consistent upon converging the timestep.

\begin{figure}
    \centering
    \includegraphics[width=10cm]{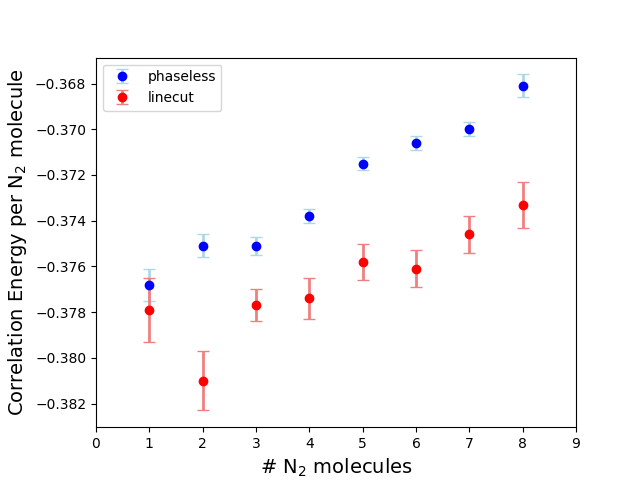} 
    \caption{Correlation energy per \ce{N2} atom for both phaseless and linecut AFQMC (cc-pVDZ) using a $0.05\ \ha$ timestep.} 
    \label{figure:size_consistency}
\end{figure} 

\newpage
\begin{spacing}{0.85}

\bibliography{References} 

\end{spacing}
\end{suppinfo}